\documentclass[11pt]{article}

\usepackage[preprint]{acl}

\usepackage{times}
\usepackage{latexsym}
\usepackage{amsmath,amssymb,amsfonts}
\usepackage{algorithmic}
\usepackage{graphicx,color}
\usepackage{textcomp}
\usepackage{hyperref}
\usepackage{graphicx}
\usepackage{hyperref}
\usepackage{amsfonts}
\usepackage{amssymb}
\usepackage{amsmath}

\usepackage{colortbl}
\usepackage[table]{xcolor}
\usepackage{booktabs}
\usepackage{multicol}
\usepackage{multirow}
\usepackage{wrapfig} 
\definecolor{blue}{RGB}{200, 235, 235}
\definecolor{red}{RGB}{255, 200, 210}
\definecolor{orange}{RGB}{255, 230, 190}
\definecolor{darkblue}{RGB}{120, 180, 180}
\definecolor{darkred}{RGB}{200, 100, 120}
\definecolor{darkorange}{RGB}{220, 150, 100}
\usepackage{algorithm,algorithmic}
\usepackage[T1]{fontenc}

\usepackage[utf8]{inputenc}

\usepackage{microtype}

\usepackage{inconsolata}

\usepackage{graphicx}

\title{Do Neural Codecs Generalize? A Controlled Study Across Unseen Languages and Non-Speech Tasks}

\author{
 \textbf{Shih-Heng Wang\textsuperscript{1}\thanks{This work was conducted during a visiting period at Carnegie Mellon University from July 2024 to November 2024.}},
 \textbf{Jiatong Shi\textsuperscript{2}},
 \textbf{Jinchuan Tian\textsuperscript{2}},
 \\
 \textbf{Haibin Wu\textsuperscript{3}},
 \textbf{Shinji Watanabe\textsuperscript{2}},
\\
 \textsuperscript{1}University of Southern California,
 \textsuperscript{2}Carnegie Mellon University,
 \textsuperscript{3}Meta
}

\begin{document}
\maketitle
\begin{abstract}
This paper investigates three crucial yet underexplored aspects of the generalization capabilities of Neural Audio Codecs (NACs): (i) Can NACs generalize to unseen languages during pre-training, (ii) Can speech-only pre-trained NACs  effectively generalize to non-speech  (e.g., environmental sounds, music, animal vocalizations, etc.) applications, and (iii) Will incorporate non-speech pre-training data boost performance on both speech and non-speech tasks? Existing studies rely on off-the-shelf NACs for comparison, yielding limited insights due to variations in NAC's implementation. In this work, we train NACs from scratch with strictly controlled configurations and carefully curated pre-trained data coverages for fair comparisons. We comprehensively evaluate NACs performance on both signal reconstruction quality and downstream application using 11 metrics. Our findings show that NACs can generalize to unseen languages during pre-training, speech-only pre-trained NACs degrade on non‐speech tasks, and including non‐speech data during pre-training improves performance on non-speech tasks while maintaining comparable performance on speech tasks.
\end{abstract}

\section{Introduction}
\label{sec:intro}
Neural Audio Codecs (NACs)~\citep{soundstream, encodec, dac, speechtokenizer, hificodec, wu2024ts3, mousavi2025discrete, arora2025landscape} have emerged as a popular signal compression technique, largely due to their strong ability to reconstruct signals after compression. Typically, NACs consist of three main components during inference: (i) an encoder, which converts raw waveforms into continuous representations; (ii) a quantizer, which transforms these continuous representations into discrete representations; and (iii) a decoder, which reconstructs the original waveform from these discrete representations. The compressed discrete tokens enable efficient transmission and  allow accurate reconstruction of the original signal. Also, these discrete representations are suitable for speech and audio language modeling tasks~\citep{valle, valle2, vallex, codecs_invest}, further underscoring their importance.

\begin{figure*}
    \centering
    \includegraphics[width=0.8\textwidth]{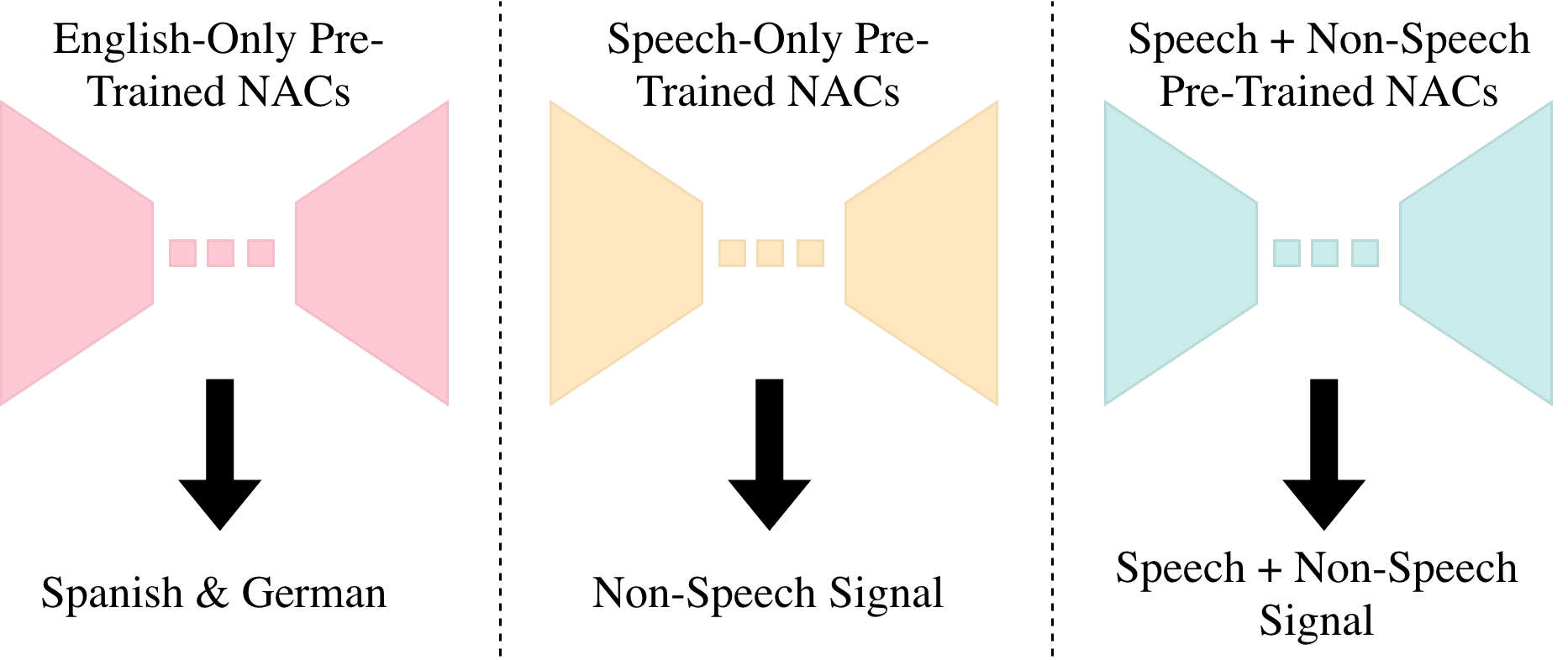}
    \caption{Overview of our three key research questions.}
    \label{fig:overview}
\end{figure*}

Despite their growing popularity, several fundamental questions about NACs remain unanswered. 
In this paper, our aim is to address three critical questions about their generalization capabilities (see Figure~\ref{fig:overview}): 

\begin{enumerate}
    \item \textbf{Can NACs generalize to unseen languages during their pre-training?}
    \item \textbf{Can speech-only pre-trained NACs effectively generalize to non-speech  (e.g., environmental sounds, music, animal vocalizations, etc.) applications?}
    \item \textbf{Will incorporate non-speech pre-training data boost performance on both speech and non-speech tasks?}
\end{enumerate}

Previous works~\citep{espnet_codec, codecsuperb, dasb, comparativestudy, usdt, survey,languagenac} focus on evaluating off-the-shelf NACs, providing limited insights into these specific questions. However, existing NACs vary in their architectures, training objectives, and pre-trained data, complicating direct comparisons and hindering detailed analysis. Clearly addressing these questions is essential for the speech community, as it guides pre-trained data curation for NACs and clarifies which NACs are appropriate for specific applications. Notably, the recent survey~\citep{survey} provides a broad overview of NACs.

Before outlining our approach, we first elaborate on the motivations behind these three questions:

\vspace{2mm}
\noindent \textbf{Q1. Can NACs generalize to unseen languages during pre-training?} Ideally, the performance of NACs should be independent to languages and subsequently fit the multilingual applications. However, existing NAC-related studies~\citep{espnet_codec, codecsuperb, wu2024codec, dasb, comparativestudy, usdt,survey} focus on single language (English) evaluation, leaving open questions about NACs' language generalization ability. Hence, our aim is to investigate whether NACs suffer from language-dependency issues. Addressing this gap is essential to determine whether NACs can be reliably applied to languages not included during their pre-training~\citep{vallex}.

\vspace{2mm}
\noindent \textbf{Q2. Can speech-only pre-trained NACs effectively generalize to non-speech applications?}
NACs are increasingly applied in non-speech scenarios, such as environmental sound detection~\citep{codecsuperb} and music generation~\citep{mucodec,pyramidcodec}. However, some existing NACs~\citep{speechtokenizer,hificodec,facodec} are trained exclusively on speech data, without exposure to non-speech signals. Given the substantial differences between these domains, such models are expected to degrade on non-speech inputs. Despite this expectation, systematic evidence is lacking. We therefore conduct experiments to empirically validate this assumption.

\vspace{2mm}
\noindent \textbf{Q3. Will incorporate non-speech pre-training data boost performance on both speech and non-speech tasks?}
Building on Q2, we investigate whether including non-speech data in the pre-training of NACs boost their performance on speech and non-speech tasks. Specifically, we aim to examine whether incorporating non-speech data can mitigate the domain gap observed in non-speech tasks.

To answer these questions, we conduct experiments under controlled conditions. Direct comparisons of existing NACs offer limited insights due to variations in their implementations. Therefore, we pre-train NACs from scratch (Sec~\ref{subsec:codec_pretrain}) with carefully curated data coverage to enable fair evaluations. Specifically, we consider three pre-trained data coverage sets (Sec~\ref{subsec:data_cover}) and evaluate the resulting NACs on both signal reconstruction quality and downstream performance (Sec~\ref{subsec:evaluation}). We utilize 11 metrics (Sec.\ref{subsec:metric}) to comprehensively assess NACs under different setups. Our results (Sec~\ref{sec:result}) show that \textbf{(i)} NACs can generalize to language unseen during their pre-training, \textbf{(ii)} speech-only NACs degrade on non-speech tasks, and \textbf{(iii)} incorporating non-speech data during pre-training improves performance on non-speech tasks while maintaining comparable performance on speech tasks.

\section{Methodology}
\label{sec:method}
Our primary goal is to conduct comparisons under controlled settings to ensure fair and consistent evaluation. To strengthen the reliability and usefulness of our findings, we make all our results fully open-source and reproducible. This section outlines our methodology First, in Sec.~\ref{subsec:codec_pretrain}, we describe our NACs pre-training process and architecture. Next, in Sec.~\ref{subsec:data_cover}, we introduce the three NACs pre-trained data coverages settings.  
Last, in Sec.~\ref{subsec:evaluation}, we explain our NACs evaluation criteria. 

\subsection{NACs Pre-Training}
\label{subsec:codec_pretrain}
Here, we will briefly go through our NACs pre-training methodology. 

\vspace{3mm}
\noindent \textbf{General framework.}  
NACs~\citep{soundstream, encodec, dac, speechtokenizer, hificodec, wu2024ts3, mousavi2025discrete, arora2025landscape} pre-training adopts a combined reconstruction and generative adversarial network (GAN) framework. During pre-training, two primary modules are involved: a generator \(\mathcal{G}\) (the codec model) and a discriminator. The generator \(\mathcal{G}\) consists of an encoder \(\mathcal{E}\), a quantizer \(\mathcal{Q}\) and codebook \(\mathcal{B}\), and a decoder \(\mathcal{D}\). After pre-training, only the generator \(\mathcal{G}\) is utilized.  

Specifically, given an input waveform \(Y = [y_1, y_2, \ldots, y_T]\), where each \(y_t \in \mathbb{R}\) and \(T\) represents its length, we first pass \(Y\) through \(\mathcal{E}\) and \(\mathcal{Q}\) to obtain the discrete representation \(\mathbf{C}\):         
\begin{equation}
  \mathbf{C} = \mathcal{Q}(\mathcal{E}(Y),\mathcal{B}).
  \label{eq:1}
\end{equation}

where \(\mathbf{C} \in \mathbb{Z}^{\hat{T} \times L}\) is a two-dimensional matrix, \(\hat{T}\) denotes the compressed sequence length, and \(L\) indicates the number of codebooks used. Next, we reconstruct the waveform \(\hat{Y}\) by applying the \(\mathcal{D}\) along with \(\mathcal{B}\) to \(\mathbf{C}\):
\begin{equation}
  \hat{Y} = \mathcal{D}(\mathbf{C}, \mathcal{B} ).
  \label{eq:2}
\end{equation}

The discriminator module \(\mathcal{D}\) comprises a short-time Fourier transform (STFT)-based discriminator and a waveform-based discriminator. These discriminators aim to distinguish between the \(Y\) and \(\hat{Y}\), thereby enhancing the perceptual quality of \(\hat{Y}\).

\vspace{2mm}
\noindent \textbf{Training loss.}
We train the model using a GAN framework, with the overall training loss comprising three components: reconstruction loss, adversarial loss, and quantization loss~\citep{soundstream, encodec, dac, speechtokenizer, hificodec, wu2024ts3, mousavi2025discrete, arora2025landscape}. Specifically, the reconstruction loss combines both time-domain and multi-scale mel-spectrogram losses. The adversarial loss includes the generator loss, discriminator loss, and feature matching loss. For the quantization loss, we apply commitment losses at both the full quantizer output and across individual quantization levels. Detailed formulations and mathematical expressions of these losses can be found in the ESPnet-Codec paper~\citep{espnet_codec}. 
We will release our pre-trained NACs and include additional implementation details in the camera-ready version.

\vspace{2mm}
\noindent \textbf{Model architecture.}  
To ensure open-source reproducibility, we employ the ESPnet-Codec toolkit~\citep{espnet_codec}, utilizing the SoundStream architecture~\citep{soundstream} with its default configurations. Specifically, \(\mathcal{E}\) and \(\mathcal{D}\) are based on SEANet~\citep{sean}, comprising primarily convolutional modules for enhanced computational efficiency. Our quantizer \(\mathcal{Q}\) adopts residual vector quantization (RVQ)~\citep{vqvae} and consists of 32 VQ modules. For efficiency, we utilize only the outputs from the first eight modules for experiments, following prior practices~\citep{valle, valle2}.  We use convolutional waveform-based and STFT-based discriminators.

\subsection{Pre-trained Data Coverages}
\label{subsec:data_cover}
To address the three questions outlined in Sec.\ref{sec:intro}, we pre-trained NACs using three distinct pre-trained data coverages: \textit{English}, \textit{Multilingual}, and \textit{Multilingual+Audio}. Specifically, the \textit{English} coverage includes only English speech data; the \textit{Multilingual} coverage includes speech data from multiple languages; and the \textit{Multilingual+Audio} coverage extends this further by incorporating both multilingual speech data and non-speech audio data. Note that the total amounts of training data differ across these configurations ( Sec.\ref{subsec:dataset}).





\subsection{Evaluation Criteria}
\label{subsec:evaluation}
We evaluated all NACs using two main criteria: \textsl{Signal Reconstruction Quality} and \textsl{Downstream Task Application}. Note that we kept all NAC parameters fixed throughout every evaluation experiment.

\vspace{2mm}
\noindent \textbf{Signal reconstruction formulation. }
The signal reconstruction capability of a NAC reflects how accurately it can recover the original waveform after discretization. We evaluate the quality of reconstructed waveforms for both speech and non-speech domains. This reconstruction process follows the formulation in Eq.~\eqref{eq:1} and Eq.~\eqref{eq:2}. Specifically, we assess the reconstruction fidelity of \(\hat{Y}\) as defined in Eq.~\eqref{eq:2}.

\vspace{2mm}
\noindent \textbf{Downstream task application. }
Beyond signal-level evaluations, it is crucial to assess how these NACs perform on real-world tasks. Therefore, we evaluate each codec in the text-to-speech (TTS) task, which is one of the most widely adopted applications ~\citep{valle,valle2,vallex} of NACs.

In TTS, each speech waveform \(Y\) is paired with a corresponding phoneme sequence \(P\) and its ground truth representation \(\mathbf{C}\) extracted from the NAC. Formally, let \( P = [p_1, p_2, \ldots, p_{M}], \quad p_t \in \mathbb{P}\), where \(M\) is the length of the phoneme sequence and \(\mathbb{P}\) is the set of all phonemes. The TTS model is represented by
\begin{equation}
  \hat{\mathbf{C}} = \textrm{TTS}(P) ,
\end{equation}
where \(\hat{\mathbf{C}}\) should approximate \(\mathbf{C}\). Then, We apply the decoder \(\mathcal{D}\) and the codebook \(\mathcal{B}\) to \(\hat{\mathbf{C}}\) (see Eq.~\ref{eq:2}) to synthesize the waveform \(\hat{Y}_{\textrm{tts}}\):
\begin{equation}
  \hat{Y}_{\textrm{tts}} = \mathcal{D}(\hat{\mathbf{C}},\mathcal{B}).
\end{equation}
The quality of \(\hat{Y}_{\textrm{tts}}\) reflects the TTS model’s performance.



\section{Experimental Setup}
\label{sec:setup}

\subsection{Dataset}
\label{subsec:dataset}

\vspace{2mm}
\noindent \textbf{Pre-trained data coverage. }
For the pre-trained data coverage mentioned in Sec~\ref{subsec:data_cover}, we used the MLS dataset\citep{mls} and AudioSet~\citep{audioset} corpora. The \textit{English} coverage consists exclusively of the English subset of MLS, while the \textit{Multilingual} coverage includes all MLS speech data across 8 languages. Finally, the \textit{Multilingual+Audio} coverage combines all MLS data with AudioSet.

\vspace{2mm}
\noindent \textbf{Speech signal reconstruction \& TTS. }
For our speech signal reconstruction and TTS experiments, we use the LJSpeech~\citep{ljspeech} and CSS10~\citep{css10} datasets, both of which are commonly adopted single-speaker TTS datasets. LJSpeech is a clean English dataset comprising approximately 24 hours of speech, while CSS10 is a multilingual dataset covering ten different languages. In this study, we only used the German and Spanish subsets of CSS10, which provide roughly 16 and 24 hours of data, respectively. All waveform files are downsampled to 16 kHz to accommodate our NACs. We report performance on the corresponding test splits. 

\vspace{2mm}
\noindent \textbf{Non-Speech Signal Reconstruction.}
For our non-speech signal reconstruction experiments, we use the AudioSet~\citep{audioset} test set, which contains 17,142 up to 10-second audio clips spanning environmental sounds, music, animal vocalizations, etc. Note that this test set is not used in our NACs pre-training. All files are downsampled to 16 kHz, and any multi-channel recordings are converted to mono-channel ones.

\subsection{Signal Reconstruction Setup}
\label{subsec:signal_setup}
We evaluate signal reconstruction performance using the pipeline described in Sec.~\ref{subsec:evaluation} on both speech and non-speech datasets (see Sec.~\ref{subsec:dataset}).  Since our NACs are trained from scratch, we provide results from established vocoder baselines as references in the speech reconstruction experiments. Specifically, these baselines convert each waveform into a mel-spectrogram and then apply a vocoder to recover the waveform. The vocoder serves as an upper bound, as it does not involve discretization. Also, we include the ground truth (GT) evaluation results as reference.

For the vocoder baseline, we employ HiFi-GAN~\citep{hifigan}, given its dominant performance among existing vocoders. We adopt the official configuration but modify the hop size to accommodate a sampling rate of 16\,kHz. We train separate HiFi-GAN models on the English, German, and Spanish datasets described in Sec.~\ref{subsec:dataset}.

\begin{table*}[h!]
    \centering
    \caption{Speech reconstruction results for NACs pre-trained with different data coverages. Bold numbers indicate the best among NACs. }
    \resizebox{\linewidth}{!}{
\begin{tabular}{l|cccc|ccc|cc|c}
    \toprule
    \multirow{3}{*}{\begin{tabular}[l]{@{}l@{}}\textbf{Data} \textbf{Coverage}\end{tabular}}
    & \multicolumn{4}{c|}{\textbf{Intrusive}}
    & \multicolumn{3}{c|}{\textbf{Non-Intrusive}}
    & \multicolumn{2}{c|}{\textbf{Perceptual}}
    & \multirow{3}{*}{\begin{tabular}[l]{@{}l@{}}\textbf{Best/} \\ \textbf{Worst}\end{tabular}}\\
    \cmidrule(lr){2-10}
    & \begin{tabular}[c]{@{}c@{}}\textbf{MCD} \\ ($\downarrow$)\end{tabular}
    & \begin{tabular}[c]{@{}c@{}}\textbf{F0-CORR} \\ ($\uparrow$)\end{tabular}
    & \begin{tabular}[c]{@{}c@{}}\textbf{PESQ} \\ ($\uparrow$)\end{tabular}
    & \begin{tabular}[c]{@{}c@{}}\textbf{S-BERT} \\ ($\uparrow$)\end{tabular}
    & \begin{tabular}[c]{@{}c@{}}\textbf{DNSMOS} \\ ($\uparrow$)\end{tabular}
    & \begin{tabular}[c]{@{}c@{}}\textbf{SHEET-SSQA} \\ ($\uparrow$)\end{tabular}
    & \begin{tabular}[c]{@{}c@{}}\textbf{UTMOS} \\ ($\uparrow$)\end{tabular}
    & \begin{tabular}[c]{@{}c@{}}\textbf{SPK-SIM} \\ ($\uparrow$)\end{tabular}
    & \begin{tabular}[c]{@{}c@{}}\textbf{WER (\%)} \\ ($\downarrow$)\end{tabular}
    &
    \\
    \midrule
    \multicolumn{11}{c}{\textbf{English}} \\
    \midrule
    \rowcolor{red}\textit{English}
        & \textbf{4.43}      
        & 0.68              
        & \textbf{2.85}     
        & \textbf{0.97}     
        & 3.14              
        & 4.21              
        & \textbf{4.05}     
        & \textbf{0.91}     
        & 3.36
        & 5 / 0                   
        \\
    \rowcolor{orange}\textit{Multilingual}
        & 4.95
        & 0.67
        & 2.63
        & 0.96
        & \textbf{3.16}     
        & 4.15
        & 3.97
        & 0.86
        & \textbf{3.21}     
        & 2 / 6
        \\
    \rowcolor{blue}\textit{Multilingual+Audio}
        & 4.63
        & \textbf{0.70}     
        & 2.77
        & 0.96
        & 3.13
        & \textbf{4.24}     
        & 3.91
        & 0.90
        & 3.38
        & 2 / 4
        \\
    \textit{Vocoder}
        & 3.76
        & 0.71
        & 3.31
        & 0.98
        & 3.15
        & 4.43
        & 4.30
        & 0.93
        & 3.33
        & --
        \\
    \textit{GT}
        & --
        & --
        & --
        & --
        & 3.17
        & 4.52
        & 4.44
        & --
        & 3.31
        & --
        \\
    \midrule
    \multicolumn{11}{c}{\textbf{Spanish}} \\
    \midrule
    \rowcolor{red}\textit{English}
        & 4.83
        & \textbf{0.44}     
        & \textbf{2.84}     
        & \textbf{0.98}     
        & \textbf{3.27}     
        & \textbf{4.14}     
        & \textbf{2.68}     
        & \textbf{0.93}     
        & \textbf{1.94}     
        & 8 / 0
        \\
    \rowcolor{orange}\textit{Multilingual}
        & 5.31
        & 0.42
        & 2.51
        & 0.91
        & \textbf{3.27}     
        & 3.75
        & 2.48
        & 0.87
        & 2.34
        & 1 / 8
        \\
    \rowcolor{blue}\textit{Multilingual+Audio}
        & \textbf{4.65}     
        & 0.43
        & 2.81
        & 0.93
        & 3.25
        & 4.07
        & 2.63
        & 0.92
        & 2.31
        & 1 / 1
        \\
    \textit{Vocoder}
        & 3.24
        & 0.49
        & 2.98
        & 0.97
        & 3.29
        & 4.44
        & 2.99
        & 0.90
        & 1.01
        & --
        \\
        \textit{GT}
        & --
        & --
        & --
        & --
        & 3.30
        & 4.51
        & 3.19
        & --
        & 2.66
        & --
        \\
    \midrule
    \multicolumn{11}{c}{\textbf{German}} \\
    \midrule
    \rowcolor{red}\textit{English}
        & \textbf{4.60}     
        & 0.63
        & 2.71
        & \textbf{0.93}     
        & \textbf{3.30}     
        & \textbf{4.11}     
        & \textbf{3.02}     
        & \textbf{0.90}     
        & \textbf{4.71}     
        & 7 / 0
        \\
    \rowcolor{orange}\textit{Multilingual}
        & 4.93
        & 0.62
        & 2.49
        & \textbf{0.93}     
        & 3.21
        & 3.93
        & 2.79
        & 0.88
        & 6.17
        & 1 / 8
        \\
    \rowcolor{blue}\textit{Multilingual+Audio}
        & 4.66
        & \textbf{0.64}     
        & \textbf{2.72}     
        & \textbf{0.93}     
        & 3.18
        & 4.06
        & 2.95
        & 0.89   
        & 4.83
        & 3 / 1
        \\
    \textit{Vocoder}
        & 3.88
        & 0.70
        & 3.08
        & 0.96
        & 3.24
        & 4.29
        & 3.31
        & 0.98
        & 3.30
        & --
        \\
     \textit{GT}
        & --
        & --
        & --
        & --
        & 3.26
        & 4.48
        & 3.60
        & --
        & 4.60
        & --
        \\   
    \bottomrule
\end{tabular}}
    \label{table:reconstruction_speech}
\end{table*}

\subsection{Downstream Application Setup}
\label{subsec:downstream_setup}
We evaluate each NAC on the TTS task (see Sec~\ref{subsec:evaluation}). For the TTS model architecture, we adopt FastSpeech2~\citep{fastspeech2}, a well-known non-autoregressive, end-to-end TTS framework. We modify FastSpeech2 to predict NAC rather than mel-spectrograms. In particular, our implementation independently predicts each of the eight codec streams. Our setup follows the NAR-TTS implementation in ESPnet-Codec, and we also source their results as our reference. Note that we do not include speaker embeddings in TTS training, as we have already verified the speaker preservation ability of NACs in the speech reconstruction experiments (see Sec.~\ref{subsec:signal_setup}).

For the baseline, we trained a continuous representation TTS model, consisting of a FastSpeech2 text-to-mel-spectrogram model and a HiFi-GAN mel-spectrogram-to-waveform vocoder. We adopt the ESPnet~\citep{espnet} implementation of FastSpeech2 with its official configuration and use the vocoder described in Sec.~\ref{subsec:signal_setup}. These continuous TTS baselines are treated as an upper bound as they do not involve any quantization process.

\subsection{Metrics}
\label{subsec:metric}
To comprehensively evaluate NACs, we use VERSA toolkit~\citep{versa}, which offers a wide range of evaluation metrics. Following ESPnet-Codec, we include the following metrics:

\vspace{2mm}
\noindent \textbf{Intrusive (Full-Reference).}
For intrusive metrics, we adopt six reference-based evaluation measures: mel cepstral distortion (MCD)\citep{mcd} for spectral distortion, F0 Pearson correlation (F0-CORR)\citep{f0corr} for pitch consistency, PESQ~\citep{pesq} for perceptual speech quality, S-BERT~\citep{sbert} for speech semantic similarity, CI-SDR~\citep{cisdr} for signal fidelity, and VISQOL~\citep{visqol} for perceptual quality modeling. All of these metrics require access to the ground-truth reference signal during evaluation.

\vspace{2mm}
\noindent \textbf{Non-Intrusive (No-Reference).}
For non-intrusive metrics, we report results for deep noise suppression mean opinion score (DNSMOS)\citep{dnsmos}, UTokyo-SaruLab’s system for the VoiceMOS Challenge (UTMOS)~\citep{utmos}, and Subjective Speech Quality Assessment in SHEET Toolkit (SHEET-SSQA)\citep{sheet}. These metrics estimate subjective quality scores without access to a reference signal, reflecting either the level of noise suppression or the perceived overall quality of the speech.

\vspace{2mm}
\noindent \textbf{Perceptual.}
For perceptual metrics, we report the word error rate (WER) using Whisper-Large~\citep{whisper}, and speaker similarity (SPK-SIM), using the speaker-embedding extractor~\footnote{{espnet/voxcelebs12\_rawnet3}} from ESPnet-SPK~\citep{spk_sim}. These metrics assess intelligibility and speaker preservation of the waveform.

We include as many metrics as possible to provide a comprehensive assessment of NACs performance across various aspects of speech quality and intelligibility.

\begin{table*}[h]
    \centering
    \caption{Non-Speech reconstruction results for NACs pre-trained with different data coverages. Bold numbers indicate the best among NACs.  }
    \resizebox{0.7\textwidth}{!}{
\begin{tabular}{l|ccc|c}
    \toprule
    \textbf{Data Coverage} & 
    \textbf{MCD ($\downarrow$)} & 
    \textbf{CI-SDR ($\uparrow$)} & 
    \textbf{VISQOL ($\uparrow$)} & 
    \textbf{Best / Worst}\\
    \midrule
    \rowcolor{red}  \textit{English} & 7.32 & -10.25 & 4.21 & 0 / 1 \\
    \rowcolor{orange}   \textit{Multilingual}   & 7.56 & -10.15 & 4.19 & 0 / 2 \\
    \rowcolor{blue}    \textit{Multilingual+Audio}   & \textbf{5.53} &  \textbf{-7.31} & \textbf{4.37} & 3 / 0 \\
    \bottomrule
\end{tabular}}
    \label{table:reconstruction_non_speech}
   \end{table*}

\begin{table*}[t]
    \centering
    \caption{TTS results for NACs pre-trained with different data coverages. Bold numbers indicate the best among NACs. }
    \resizebox{0.9\textwidth}{!}{
\begin{tabular}{l|c|ccc|c}
    \toprule
    \textbf{Data Coverage} &
    \textbf{WER (\%) ($\downarrow$)} & 
    \textbf{UTMOS ($\uparrow$)} & 
    \textbf{DNSMOS ($\uparrow$)} & 
    \textbf{SHEET-SSQA ($\uparrow$)} &
    \textbf{Best / Worst} \\
    
    \midrule
    \multicolumn{6}{c}{\textbf{English}} \\
    \midrule
    \rowcolor{red}\textit{English}             
      & \textbf{4.33}   
      & 2.36 
      & 2.36
      & 3.06 
      & 1 / 3 \\
    \rowcolor{orange}\textit{Multilingual}         
      & 4.83
      & \textbf{2.84}   
      & \textbf{2.72}   
      & \textbf{3.26}   
      & 3 / 1
      \\
    \rowcolor{blue}\textit{Multilingual+Audio}   
      & 4.71
      & 2.47
      & 2.44
      & 3.11 
      & 0 / 0 \\
    \textit{Continuous TTS}        
      & 3.63
      & 4.23
      & 3.06
      & 4.15 
      & -- \\
    \textit{ESPnet-Codec~\cite{espnet_codec}} 
      & 4.70
      & 1.84
      & --
      & -- 
      & -- \\
    \midrule
    \multicolumn{6}{c}{\textbf{Spanish}} \\
    \midrule
    \rowcolor{red}\textit{English}             
      & \textbf{4.04}   
      & 1.75
      & 2.95
      & 2.65 
      & 1 / 1 \\
    \rowcolor{orange}\textit{Multilingual}         
      & 6.89
      & \textbf{1.95}   
      & \textbf{3.07}   
      & \textbf{2.88}   
      & 3 / 1\\
    \rowcolor{blue}\textit{Multilingual+Audio}   
      & 6.30
      & 1.64
      & 2.77
      & 2.67 
      & 0 / 2\\
    \textit{Continuous TTS}        
      & 4.72
      & 3.40
      & 3.18
      & 4.58 
      & -- \\
    \midrule
    \multicolumn{6}{c}{\textbf{German}} \\
    \midrule
    \rowcolor{red}\textit{English}             
      & \textbf{13.30}  
      & \textbf{1.51}   
      & 2.16
      & 2.29 
      & 2 / 0\\
    \rowcolor{orange}\textit{Multilingual}         
      & 16.13
      & 1.49
      & \textbf{2.36}   
      & 2.17 
      & 1 / 3 \\
    \rowcolor{blue}\textit{Multilingual+Audio}   
      & 13.55
      & \textbf{1.51}   
      & 2.05
      & \textbf{2.38}   
      &  2 / 1\\
    \textit{Continuous TTS}        
      & 11.21
      & 3.35
      & 2.60
      & 4.04 
      & -- \\
    \bottomrule
\end{tabular}}
    \label{table:tts}
\end{table*}
\section{Results \& Analysis}
\label{sec:result}
\subsection{Result}
\label{subsec:result}

\vspace{2mm}
\noindent \textbf{Signal reconstruction - Speech. }
In Table~\ref{table:reconstruction_speech}, we present the signal reconstruction results on the speech dataset test sets (see Sec~\ref{subsec:dataset}). 
The table is split into three sections, each corresponding to the language of the data used for evaluation. 
Each row is color-coded to reflect the NAC’s pre-training coverage: 
red for \textit{English}, orange for \textit{Multilingual}, blue for \textit{Multilingual+Audio}, with \textit{vocoder and GT} shown without any color. To facilitate comparisons, we compute a ``Best / Worst'' for each NAC, representing how many metrics it performs the best / worst in each language. Results across different languages are not comparable.

\vspace{1.5mm}
Table~\ref{table:reconstruction_speech} indicates that the NAC pre-trained with \textit{English} data coverage generally outperforms those pre-trained with other coverages in speech reconstruction tasks.  
Specifically, the NAC pre-trained on \textit{English} data coverages achieves the best performance in 5, 8, and 7 metrics (out of 9) for English, Spanish, and German, respectively. 

\vspace{2.5mm}
\noindent \textbf{Signal reconstruction - Non-Speech. }
In Table~\ref{table:reconstruction_non_speech}, we present signal reconstruction results on the AudioSet~\citep{audioset} test sets (see Sec.~\ref{subsec:dataset}). Following ESPnet-Codec~\citep{espnet_codec}, we report three signal-level metrics: MCD, CI-SDR, and VISQOL.

From Table~\ref{table:reconstruction_non_speech}, we observe the NAC pre-trained with \textit{Multilingual+Audio} coverage outperforms those pre-trained without non-speech data across 3 metrics. Note that CI-SDR values may be negative for poor reconstructions; however, relative improvements (e.g., from -10 dB to -7 dB) still indicate meaningful quality gains.

\vspace{2mm}
\noindent \textbf{Downstream Application.}  
In Table~\ref{table:tts}, we present TTS results for each NAC. The reported metrics include WER, which evaluates the intelligibility, and three non-intrusive metrics that assess the quality of the synthesized waveform.

In terms of WER, the NAC pre-trained with \textit{English} data coverage achieves the best results, followed by the \textit{Multilingual+Audio}. For the non-intrusive quality metrics, the NAC pre-trained on \textit{Multilingual} data coverage generally performs best. Notably, the NAC pre-trained on \textit{English} ranks second in Spanish and German, unseen languages during its pre-training. Note that the UTMOS scores for our results remain relatively low, especially for Spanish and German . We attribute this to the inherent limitations of training TTS models with NACs on small datasets, as well as the limited cross-lingual generalization capability of the UTMOS model. Prior studies~\citep{espnet_codec, dasb} have reported similarly low UTMOS scores, indicating that this trend is consistent with existing findings rather than an anomaly.
\subsection{Analysis}

\vspace{1.5mm}
\noindent \textbf{Q1.}
From Table~\ref{table:reconstruction_speech} and Table~\ref{table:tts}, NACs demonstrate strong generalization to languages unseen during pre-training. For both Spanish and German, the NAC pre-trained on \textit{English} data (Best / Worst: 15 / 0 in Table~\ref{table:reconstruction_speech}; 3 / 1 in Table~\ref{table:tts}) achieves generally more best counts than the \textit{Multilingual} one (Best / Worst: 2 / 16 in Table~\ref{table:reconstruction_speech}; 4 / 4 in Table~\ref{table:tts}). This contrasts with speech self-supervised learning (SSL) models~\citep{mlsuperb, mlsuperb2, hokkien, sslstan}, which typically degrade on unseen languages during pre-training. These results highlight the language robustness of NACs and their potential for broader linguistic applications.

\vspace{2mm}
\noindent \textbf{Q2.}
As shown in Table~\ref{table:reconstruction_non_speech}, speech-only pre-trained NACs (Best / Worst: 0 / 2) perform worse than speech and non-speech pre-trained NACs (Best / Worst: 3 / 0). This aligns with expectations, as speech and non-speech signals differ substantially at the signal level.

\vspace{2mm}
\noindent \textbf{Q3.}
As shown in Table~\ref{table:reconstruction_speech}, Table~\ref{table:reconstruction_non_speech}, and Table~\ref{table:tts}, incorporating non-speech data during pre-training improves performance on non-speech tasks (see Q2 above) while maintaining acceptable performance on speech tasks (Best / Worst: 6 / 6 in Table~\ref{table:reconstruction_speech}; 2 / 3 in Table~\ref{table:tts}). This finding is somewhat unexpected, as we initially hypothesized that including non-speech data might introduce trade-offs. Based on these results, we recommend including non-speech data in NAC pre-training to enhance generalization across speech and non-speech domains.

\section{Conclusion}
This paper addresses three crucial yet underexamined questions regarding neural audio codecs (NACs). First, we pre-train NACs using aligned model configurations and three carefully curated pre-training data coverages to enable fair comparisons. We then evaluate their performance on signal reconstruction and downstream tasks using 11 complementary metrics for a comprehensive assessment. Our results show that (i) NACs can generalize to languages unseen during pre-training, (ii) NACs pre-trained exclusively on speech data perform poorly on non-speech tasks, and (iii) incorporating non-speech data during pre-training improves performance on non-speech tasks while maintaining comparable performance on speech tasks. These findings suggest that NACs can be reliably applied to multilingual settings, that non-speech-aware pre-training is essential for non-speech applications, and that joint pre-training on both speech and non-speech data offers the most robust overall performance.

\section{Limitations}

\vspace{2mm}
\noindent \textbf{NAC architecture.}
In this work, we adopt the SoundStream architecture~\citep{soundstream} with its default configurations. In future work, we plan to extend our study to other NAC architectures, such as EnCodec~\citep{encodec}, DAC~\citep{dac}, and HiFi-Codec~\citep{hificodec}, to further validate the generality of our findings.

\vspace{2mm}
\noindent \textbf{Evaluation criteria.}
We evaluate NACs from two perspectives. For signal reconstruction, we consider both speech and non-speech signals. For downstream applications, however, we focus solely on TTS, which is inherently a speech-only task. In future work, we plan to include additional non-speech downstream tasks, such as music generation and environmental sounds related tasks, to provide a more comprehensive evaluation. 

\vspace{2mm}
\noindent \textbf{TTS architectures.} We plan to extend our TTS experiments to more powerful autoregressive models, such as VALL-E~\citep{valle, valle2, vallex}. While these models are computationally more demanding, they may offer stronger generative capabilities and provide further insights into the interaction between NACs and advanced TTS architectures.

\bibliography{custom}




\end{document}